\documentclass[12pt,preprint,showpacs,preprintnumbers,amsmath,amssymb]{revtex4}

\usepackage{graphicx}

\begin{document}

\title{Generation of a decoherence-free entangled state using a radio frequency dressed state}

\author{Atsushi~Noguchi$^{1}$}
\email[]{noguchi@qe.ee.es.osaka-u.ac.jp}
\author{Shinsuke~Haze$^{1}$}
\author{Kenji~Toyoda$^{1}$}
\author{Shinji~Urabe$^{1}$}
\affiliation{%
$^{1}$Graduate School of Engineering Science, Osaka University, 1-3 Machikaneyama, Toyonaka, Osaka, Japan}

\date{\today}
             
\begin{abstract}
We propose the generation of entangled states with trapped calcium ions using a combination of an rf dressed state and a spin dependent force.
Using this method, a decoherence-free entangled state of rf qubits can be directly generated and ideally its fidelity is close to unity.
We demonstrate an rf entangled state with a fidelity of 0.68 $\pm 0.08$, which has a coherence time of more than $200\mathrm{ms}$ by virtue of the fact that it is an eigenstate with energy gaps between adjacent levels.
Using the same technique, we also produce a qutrit-qutrit entangled state with a fidelity of $0.77\pm 0.09$, which exceeds the threshold value for separability of 2/3.
\end{abstract}

\maketitle

Quantum information processing (QIP) using atomic ions trapped in a linear Paul trap has been widely investigated and has been successfully achieved\cite{1,2,3}, because collective motion can be used to achieve two-qubit operations\cite{3,4}. In addition, recent experiments on scalability have demonstrated the entanglement of 14 ions\cite{5} and quantum simulation of 9 ions\cite{6}, based on a spin dependent force, a Molmer-Sorensen (MS) interaction\cite{7,8}.
For an rf qubit, for which the energy separation between the two states corresponds to radio frequency, excitation of collective motion is difficult without special techniques\cite{9} due to the extremely low Lamb-Dicke parameters.
However, to achieve a scalable quantum computer it is advantageous to manipulate qubits with rf or microwave magnetic fields\cite{9,10}. 
Thus, the direct generation of entanglement of rf qubits is one of building blocks of QIP with trapped ions.

Another requirement for quantum computers is a long coherence time for qubits that are constantly exposed to disturbances from the environment such as stray magnetic fields\cite{11}. A noisy environment induces unwanted disturbances to quantum states and leads to decoherence.
From this point of view, a decoherence free subspace (DFS) is well known to prolong the coherence time\cite{12,13}.  In particular, a DFS using a dressed state has been recently demonstrated for a single qubit\cite{10}. 
Such a "dressed DFS" has the advantage that quantum states are protected even in the presence of a stray magnetic field gradient, which often limits the coherence time in a traditional DFS\cite{12}.
In the dressed DFS, quantum states which might be coupled to each other by magnetic fields are energetically separated.
If qubits are globally  exposed to an external field (dressing field), a Dicke ladder is constructed.
In particular, states on the Dicke ladder whose projections of total angular momentum are zero are not perturbed by the dressing field. In the case of two ions, these states can be used to construct a logical qubit.
Moreover, protected large-scale entangled states also exist when a large number of ions are globally  subjected to a dressing field.

In this paper, we demonstrate  the combination of a dressed state of rf qubits and an MS interaction.
This method is based on the concept of inducing inter-atomic correlation using an ancillary state rather than the qubit states.
The rf dressed states produce a dressed DFS and one of these states is connected to the ancillary state through the MS interaction.
Thus, using this combination, we can directly generate an entangled state of rf qubits, which is a Dicke state that spans the dressed DFS.

\begin{figure}[b]
   \includegraphics[width=7.5cm,angle=-90]{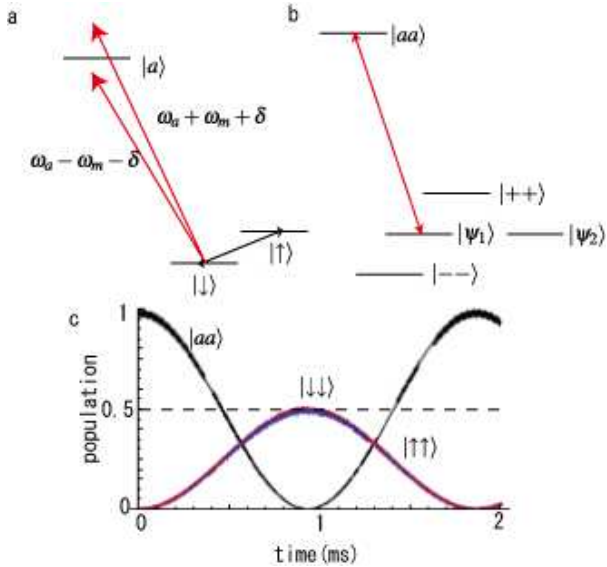}
\caption{(a) Energy-level diagram for a robust entangled state. $\omega _a$ is the frequency of the transition$\lvert a\rangle\leftrightarrow\lvert\downarrow\rangle$, $\omega _m$ is the normal mode frequency and $\delta$ is the amount of detuning from the sideband transition.
(b) Energy-level diagram for two ions case upon the dressed state picture. 
(c) Numerically calculated populations based on the Hamiltonian $\hat{H}_I$. The three curves correspond to the population of the $\lvert aa\rangle$, $\lvert\downarrow\downarrow\rangle$ and $\lvert\uparrow\uparrow\rangle$ states.  The calculation parameters are $\Omega _1\ =\ 2\pi\ \times\ 300\ \mathrm{kHz}$, $\eta\Omega _2\ =\ 2\pi\ \times\ 20\ \mathrm{kHz}$ and $\delta\ =\ 2\pi\ \times\ 40\ \mathrm{kHz}$. 
}
\label{model}
\end{figure}

We first consider the situation depicted in Fig. 1a. 
There are three (meta)stable states, namely an ancillary state, $\lvert a\rangle$, and one qubit, $\{\lvert\downarrow\rangle,\lvert\uparrow\rangle\}$.
The interaction among these states can be expressed by the Hamiltonian
\begin{eqnarray*}
\hat{H}_I &=&\hat{H}_1+\hat{H}_2\\
\hat{H}_1&=&\sum _{i}\frac{\hbar\Omega _1}{2} (\hat{\sigma}^+_{1,i}+\mathrm{h.c.})\\
\hat{H}_2&=&\sum _{i}\frac{\hbar\eta \Omega _2}{2} (\hat{a}\hat{\sigma} ^+_{2,i} e^{-i \delta t} +\hat{a}^\dagger \hat{\sigma} ^+_{2,i} e^{-i\delta t} +\mathrm{h.c.})
\end{eqnarray*}
where $\hat{a}$ and $\hat{a}^\dagger$ are the annihilation and creation operators of the motional mode, respectively, and $\hat{\sigma}^+_{1,i}$ and $\hat{\sigma}^-_{1,i}$ ($\hat{\sigma}^+_{2,i}$ and $\hat{\sigma}^-_{2,i}$) are the spin flip operators between the $\{\lvert\downarrow\rangle , \lvert\uparrow\rangle\}$ ($\{\lvert a\rangle , \lvert\downarrow\rangle\}$) states of the i-th ion. 
The first term ($\hat{H}_1$) gives rise to dressed states of the qubit,  expressed as $\lvert +\rangle=(\lvert\uparrow\rangle+\lvert\downarrow\rangle)/\sqrt{2}$ and $\lvert -\rangle=(\lvert\uparrow\rangle-\lvert\downarrow\rangle)/\sqrt{2}$. $\Omega _1$ is the Rabi frequency of the transition between the qubit states.
The second term in the Hamiltonian ($\hat{H}_2$) corresponds to an MS type interaction\cite{7}. $\eta \Omega _2$ is the Rabi frequency of the sideband transitions and $\delta$ is the amount of detuning.
We now consider the case of two ions (Fig. 1b). 
If the first term in the Hamiltonian is much larger than the second, the Hamiltonian has three eigenspaces $\{\lvert ++\rangle\}, \{\lvert\psi _1\rangle , \lvert\psi _2\rangle\}, and \{\lvert --\rangle\}$, whose eigenvalues are $2\hbar \Omega _1$, 0, and $-2\hbar\Omega _1$ respectively, where$\lvert \psi _1\rangle\ =(\lvert +-\rangle+\lvert -+\rangle )/\sqrt{2}\ (=(\lvert\uparrow\uparrow\rangle -\lvert\downarrow\downarrow\rangle )/\sqrt{2})$ and $\lvert\psi_2\rangle\ =(\lvert +-\rangle-\lvert -+\rangle )/\sqrt{2}\ (=(\lvert\uparrow\downarrow\rangle -\lvert\downarrow\uparrow\rangle )/\sqrt{2})$.
The two states $\lvert\psi _1\rangle and \lvert \psi _2\rangle$ span a "dark space", in which no states are affected by the Hamiltonian $\hat{H}_1$. 

On the other hand, the MS interaction selectively couples one dark state $\lvert\psi _1\rangle$ in the dark space with $\lvert aa\rangle$, because the MS interaction can be considered to be a transition between the $\lvert aa\rangle$ and $\lvert \downarrow\downarrow\rangle$ states, and the other dressed states either do not contain $\lvert \downarrow\downarrow\rangle$ or have energy gaps due to $\hat{H}_1$. So the full Hamiltonian describes a transition between the $\lvert aa\rangle$ and $\lvert \psi _1\rangle$ states if the first term is much larger than the second, i.e., $\Omega _1\gg\eta \Omega _2$.
Please note that although the $\lvert \psi _1\rangle$ state is generally sensitive to stray magnetic fields, it is robust against fluctuating magnetic fields if we apply only the first term of the Hamiltonian. This state is a dark state of the Hamiltonian and the $\{\lvert ++\rangle and \lvert --\rangle\}$ states are energetically separated, so that it is less likely for the dark state to be coupled to these states by stray magnetic fields\cite{10}.
Moreover, as the total angular momentum of the $\lvert \psi _2\rangle$ state is 0, the global operation does not couple the $\lvert \psi _1\rangle$ and $\lvert \psi _2\rangle$ states. 
Fig. 1c shows numerical results for the state populations based on the Hamiltonian $\hat{H}_I$.  The initial state is $\lvert aa\rangle$ and the populations of the three states $\lvert aa\rangle, \lvert \downarrow\downarrow\rangle \text{ and }\lvert \uparrow\uparrow\rangle$ are depicted as solid curves. 
The initial state, $\lvert aa\rangle$, evolves to the entangled state $\lvert \psi _1\rangle$.
The parameters used in the calculation are shown in the caption of Fig. 1c.
We consider an MS interaction in the fast regime\cite{4} where a significant amount of collective motion is excited. However, the first term, $\hat{H}_1$, reduces the excitation of the motional mode in the case of $\Omega _1>\eta \Omega _2$, because the intermediate states of the MS interaction ($\{\lvert a\downarrow\rangle , \lvert \downarrow a\rangle\}$) become dressed states of $\hat{H}_1$ and the effective detuning for the MS interaction becomes large.
The $\lvert \psi _1\rangle$ state is then generated with little motional excitation.
Although the characteristics of the MS interaction change from the case without the dressing field, we have numerically verified that the excitation of the transition $\lvert aa\rangle \leftrightarrow\lvert \psi _1\rangle$ can be achieved using the Hamiltonian $\hat{H}_I$ and ideally its fidelity is unity in the case of two ions. 
Moreover, because we consider a Hilbert space with three quantum states, we can construct a qutrit\cite{14}  consisting of $\{\lvert +1\rangle ,\lvert 0\rangle ,\lvert -1\rangle\}=\{\lvert a\rangle ,\lvert \downarrow\rangle ,\lvert \uparrow\rangle\}$, which can be adapted to specific purposes, such as enabling efficient use of communication channels in quantum cryptography\cite{15}.
When the population of all three states becomes equal (e.g., at 0.57 ms in Fig. 1c) an entangled qutrit-qutrit state is generated.

\begin{figure}[t]
   \includegraphics[width=7.5cm,angle=-90]{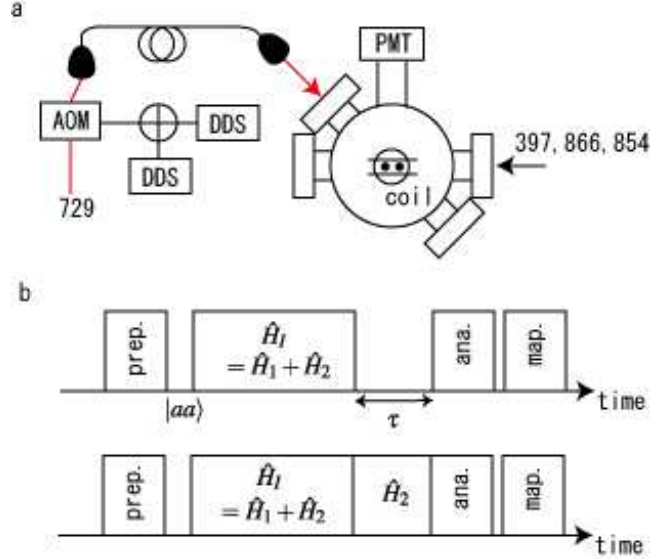}
\caption{(a) Experimental setup. Two ions trapped in a linear Paul trap are irradiated with two 729 beams, whose wavelengths are shifted using an acousto-optic modulator driven at two frequencies by two phase-locked direct digital synthesizers. An rf magnetic field is induced by a coil. (b) Pulse sequences for experiments. }
\label{setup}
\end{figure}

In our experimental setup, which is depicted in Fig. 2a, two $^{40}\mathrm{Ca}^+$ ions are confined along the axis $(\hat{z})$ of a linear Paul trap whose normal mode frequencies for a single $^{40}\mathrm{Ca}^+$ ion are $\{ \omega _x, \omega _y,\omega _z \}$ = $2\pi\ \times$ $\{ 2.9, 2.9, 1.0\}$ MHz\cite{16}.
The qubit$(\lvert \downarrow\rangle ,\ \lvert \uparrow\rangle )$ is constructed based on the electronic spin of the ground state $^2\mathrm{S}_{1/2}\ (m=\pm \frac{1}{2})$, and the ancillary state $(\lvert a\rangle)$ is one of the excited states $^2\mathrm{D}_{5/2}\ (m^\prime =-\frac{5}{2})$.
The two qubit states are coupled using an rf magnetic field induced by a coil located below the trap\cite{17}, which gives rise to the rf dressed states $\lvert \psi _1\rangle and \lvert \psi _2\rangle$. 
The $\lvert \downarrow\rangle$ state is coupled to the $\lvert a\rangle$ state by a quadrupole transition $^2\mathrm{S}_{1/2} \leftrightarrow ^2\mathrm{D}_{5/2}$ excited by a 729-nm laser beam with a narrow linewidth $(\sim 2\pi\ \times \ 300\ \mathrm{Hz})$. 
Because of its low heating rate, we utilize a stretch mode with a secular frequency of $\omega _{str}\ =\ 2\pi\times$ $1.75$ MHz.

The two ions are first subjected to Doppler cooling and placed in the ground states ($\langle n_{str}\rangle\sim 0.09$) of all motional modes through sideband cooling cycles.
For the preliminary experiment, we irradiate the ions with two laser beams with different wavelengths whose detunings from the optical carrier transition are $\pm\omega _{str}\pm\delta$, where $\delta = 2\pi \times 17.6\mathrm{kHz}$. The Rabi frequencies of these two beams are $2\pi\ \times\ 8.8\mathrm{KHz}$. 
The two beams induce an MS interaction between the ions and generate an entangled state $(\lvert aa\rangle +\lvert \downarrow\downarrow\rangle )/\sqrt{2}$, which has a fidelity of 0.91 in our experiment. 
In this experiment we set  the detuning to $\delta =2\eta\Omega _2$ in order to quicken the time evolution of the interaction. 
Fig. 2b shows the pulse sequences used in the experiment.
After cooling, both ions are polarized to the $\lvert \downarrow\rangle$ state and are irradiated with a $\pi$ pulse of the optical carrier transition to place them in the initial state, $\lvert aa\rangle$.
An interaction with $\hat{H}_I$ is then induced using the laser beams and the rf magnetic field.
After the interaction we wait for a time ($\tau$) without or with the rf magnetic field in order to measure the coherence time. 
At the end of the sequence these states must be mapped to the excited states $^2\mathrm{D}_{5/2}$ for the projection measurement.

\begin{figure}[t]
   \includegraphics[width=7.5cm,angle=-90]{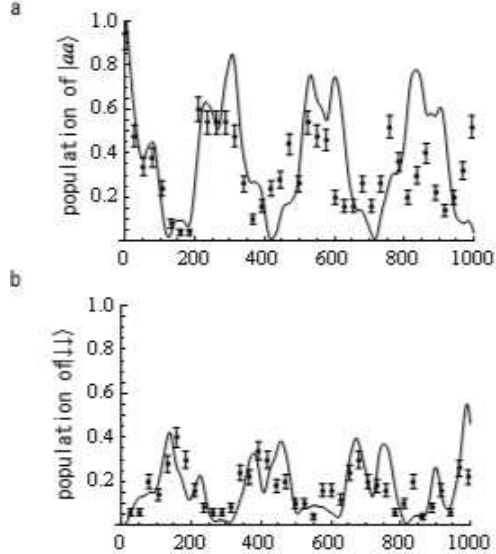}
\caption{The evolution of the state populations when the Hamiltonian $\hat{H}_I$ is applied.  (a) without mapping and (b) with mapping. The vertical axis is the population of each state and the horizontal axis is the interaction time.  The solid curves show the results of numerical calculations using the experimental parameters (without any fitting).}
\label{time}
\end{figure}

Fig. 3 shows the evolution of the state populations when the Hamiltonian $\hat{H}_I$ is applied. The ions are irradiated with two laser beams with Rabi frequencies of $2\pi\ \times\ 8.8\mathrm{kHz}$ and detunings of $\delta\ =\pm 2\pi\ \times 17.6\ \mathrm{kHz}$ from the sideband transitions, and induce an rf magnetic field whose Rabi frequency is $2\pi\ \times\ 10.5 \mathrm{kHz}$.
The experimental results for the time evolution of the state populations are depicted by the data points in Fig. 3a and 3b for the case without and with a mapping pulse, respectively. The solid curves are theoretical results without any fitting, and are seen to be in good agreement with the experimental data.
When the population of $\lvert \downarrow\downarrow\rangle$ is at a maximum (t=157 $\mathrm{\mu s}$), an entangled state of the rf qubits will be realized.
At this point, to confirm the presence of entanglement, the diagonal elements of the rf qubit are measured using mapping pulses. 
The measured populations in the $\{ \lvert \uparrow\uparrow\rangle ,\lvert \downarrow\downarrow\rangle \}$ states are $\{0.34\pm 0.04, 0.40\pm 0.04 \}$.

To estimate the fidelity, a $\pi /2$ rf pulse (whose phase is $\theta$) is applied before the mapping pulses and the non-diagonal elements of the rf qubits are measured. 
Fig. 4a shows the parity oscillation whose amplitude is determined to be $0.62 \pm 0.08$ by fitting.
The fidelity of this state is estimated to be $0.68\pm 0.08$, which is above the threshold of 0.5 for separability even if the error bars are considered.
There are two reason why the fidelity is less than unity.
The first is imperfection of the MS interaction due to differences between the Rabi frequencies of the ions and the amount of detuning of the sideband beams.
The second is the low Rabi frequency of the rf magnetic field, which will be discussed in detail later.

\begin{figure}[t]
   \includegraphics[width=7.5cm]{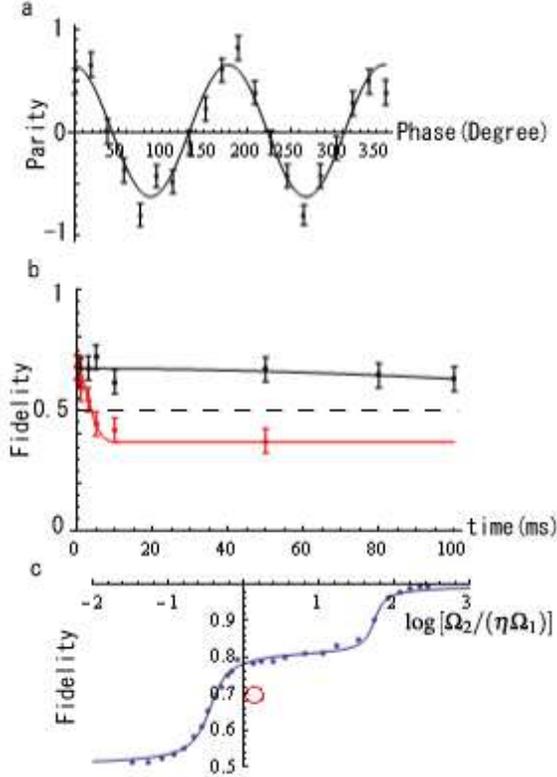}
\caption{(a) Parity oscillation to measure non-diagonal elements of rf qubit. 
(b) Coherence time of $\lvert \psi_1\rangle$ state. The horizontal axis is the time $\tau$ from Fig. 2b. Black and red dots and curves represent the case with and without an applied rf magnetic field (Hamiltonian $\hat{H}_2$), respectively. The error bars correspond to the statistical standard error.
(c) Fidelity vs. Rabi frequency of rf field determined by numerical calculation. 
The blue dots are the numerical results and the blue solid curve is a visual guide. The red circle shows the result obtained in the present experiment.}
\label{cohe}
\end{figure}

Fig. 4b shows the coherence time of the generated state. Because the $\lvert \psi _1\rangle\ (=(\lvert \uparrow\uparrow\rangle -\lvert \downarrow\downarrow\rangle )/\sqrt{2})$ state is sensitive to static magnetic fields, the coherence time is limited to a few milliseconds in our experiments (the red dots in Fig 4b).
However, if we apply an rf magnetic field after the interaction, the $\lvert \psi _1\rangle$ state becomes a dark state and the coherence time increases.
The black dots in Fig. 4b show the fidelity as a function of the time interval $\tau$ between the interaction and analysis pulses (see Fig. 2b), for the case when the rf magnetic field applied.
By fitting to the experimental data, the coherence times are determined to be $200 \mathrm{ms}^{+200 \mathrm{ms}}_{-60\mathrm{ms}}$  and $(4.3\mathrm{ms}\pm 0.6\mathrm{ms})$, with and without the rf magnetic field, respectively. 
For intervals longer than 100 ms, we can not measure the coherence time for technical reasons, one of which is instability of a coulomb crystal. 
Nevertheless, when the rf magnetic field is used, the fidelity is seen to decrease very little up to a period of 100 ms and the coherence time is increased by about two orders of magnitude. 
Using this interaction, an entangled state can also be produced using a qutrit of two ions ($\lvert \Psi\rangle\ =\ e^{i\theta}\lvert aa\rangle +\lvert \downarrow\downarrow\rangle -\lvert \uparrow\uparrow\rangle$).
This "GHZ-like qutrit entangled state" can be evaluated in a similar way to GHZ states, without the need for full quantum tomography.
The fidelity of this maximally entangled state can be calculated by
\[
F_3\ =\ \sum _i \frac{1}{2}(D_i +A_i /2)
\]
where $D_i$ are the diagonal elements for the $\{\lvert aa\rangle , \lvert \downarrow\downarrow\rangle , \lvert \uparrow\uparrow\rangle\}$ states and $A_i$ are the amplitudes of the three parity oscillations among the $\{ \lvert a\rangle ,\lvert \downarrow\rangle\}$,$\{ \lvert a\rangle ,\lvert \uparrow\rangle\}$ and $\{ \lvert \uparrow\rangle ,\lvert \downarrow\rangle\}$ states.
Applying the same technique used in the qubit entanglement experiment, the $\lvert \Psi\rangle$ state is generated and the density matrix elements are measured to be $\{D_1,D_2,D_3, A_1, A_2, A_3 \}$=$\{0.33,0.3,0.3,0.54,0.42,0.27\}$. 
The results show a fidelity of $0.77\pm 0.09$, which is larger than 2/3 even considering the error bars.
There are two reasons for the higher fidelity of qutrit-qutrit entanglement. 
First, the interaction time for generating entangled states is shorter in the qutrit case. The coherence of the overall system is limited by the linewidth of the 729-nm laser beam.
Second, in the qutrit case, the fidelity of the diagonal elements is large because of the contribution of the population of the $\lvert aa\rangle$ state.

We check the validity of the concept of rf dressed states.
This concept only becomes invalid when the interaction $\hat{H}_2$ is larger than $\hat{H}_1$. 
We therefore numerically calculate the fidelity as a function of the Rabi frequency of the rf magnetic field, and the results are shown in Fig. 4c.
Here, the blue dots show the calculation results and the blue solid curve is a visual guide.
A stronger rf magnetic field is more effective at generating an entangled state.
From the figure, it can be seen that there are two steps at $\Omega _1/(\eta \Omega _2)\sim 6 \text{ and }0.6$. In order to produce an entangled state with a fidelity above 0.99, an rf Rabi frequency larger than the optical Rabi frequency by one order of magnitude is necessary.
The red circle in Fig. 4c shows the fidelity and Rabi frequency in the present experiment.
The low experimental fidelity of the entangled state may be explained by the fact that even if all other conditions are ideal, the fidelity is limited to 0.8 due to the low Rabi frequency of the rf magnetic field compared to the optical Rabi frequency. 
The experimental fidelity ($0.68\pm 0.08$) is consistent with the value of $\sim 0.66$ obtained from $0.8\times 0.91^2$, where the first term represents the limitation imposed by the low rf Rabi frequency and the value of 0.91 is a factor representing the imperfect MS interaction. 

In summary, we combined an MS interaction and a dressed state and directly generated an entangled state of an rf qubit in a dressed DFS.
The fidelity of this entangled state was $0.68$ and by applying an rf magnetic field the coherence time of this state was prolonged from $4.3\ \mathrm{ms}$ to more than $200\ \mathrm{ms}$. 
If a strong enough rf magnetic field with a Rabi frequency much larger than that of the optical field is used, this method can generate an rf entangled state with a high fidelity.
This dressed DFS can be generalized to a large number of ions and the coherence of the large-scale entangled states can be protected in the dressing field.
Using this method, we also demonstrated the entanglement of qutrits, and achieved a fidelity of $0.77$.

{\large{Acknowledgements}}

This work was supported by MEXT Kakenhi "Quantum Cybernetics" Project and the JSPS through its FIRST Program.
One of the authors (N. A.) was supported in part by the Japan Society for the Promotion of Science.

{\large{Author Contribution}}

A.N. conceived the experiment and analysed the data; S.U., K.T. and S.H. contributed to the experimental set-up; and all authors co-wrote the paper.


\begin{thebibliography}{99} 
\bibitem{1}Quantum State Engineering on an Optical Transition and Decoherence in a Paul Trap. Roos, Ch. et al.  \textit{Phys. Rev. Lett.}  \textbf{83}, 4713  (1999).

\bibitem{2}Realization of the Cirac-Zoller controlled-NOT quantum gate. Schmidt-Kaler, F. et al. \textit{Nature} \textbf{422}, 408 (2003).

\bibitem{3}Control and Measurement of Three-Qubit Entanglement States. Roos, C. F. et al. \textit{Science} \textbf{304}, 1478 (2004).

\bibitem{4}Experimental entanglement of four particles. Sackett, C. A. et al. \textit{Nature} \textbf{404}, 256 (2000).

\bibitem{5}14-Qubit Entanglement: Creation and Coherence. Monz, T. et al. \textit{Phys. Rev. Lett.} \textbf{106}, 130506 (2011).

\bibitem{6}Onset of a quantum phse transition with a trapped ion quantum simulator. Islam, R. et al. \textit{Nature Communications} \textbf{2}, 377 (2011).

\bibitem{7}Multiple Entanglement of Hot Trapped Ions. Molmer, K. \& Sorensen, A. \textit{Phys. Rev. Lett.} \textbf{82}, 1835 (1999).

\bibitem{8}Quantum Computation with Ions in Thermal Motion. Sorensen, A. \& Molmer, K. \textit{Phys. Rev. Lett.} \textbf{82}, 1971 (1999).

\bibitem{9}Microwave quantum logic gates for trapped ions. Ospellkaus, C. et al. \textit{Nature} \textbf{476}, 181 (2011).

\bibitem{10}Quantum gates and memory using microwave-dressed states. Timoney, N. et al. \textit{Nature} \textbf{476}, 185 (2011).

\bibitem{11}Wiring up quantum systems. Schoelkopf, R. J. \& Girvin, S. M. \textit{Nature} \textbf{451}, 664 (2008). 

\bibitem{12}Long-Lived Qubit Memory Using Atomic Ions. Langer, C. et al. \textit{Phys. Rev. Lett.} \textbf{95}, 060502 (2005).

\bibitem{13}A Decoherence-Free Quantum Memory Using Trapped ions. Kielpinski, D. et al. \textit{Science} \textbf{291}, 1013 (2001)

\bibitem{14}Measuring Qutrit-Qutrit Entanglement of Orbital Angular Momentum State of an Atomic Ensenble and a Photon. Inoue, R. et al. \textit{Phys. Rev. Lett.} \textbf{103}, 110503 (2009).

\bibitem{15}Quantum Key Distribution with High-Order Alphabets Using Spatially Encoded Qudits. Walborn, S. P, Lemelle, D. S., Almeida, M. P. \& Souto Ribeiro, P. H. \textit{Phys. Rev. Lett.} \textbf{96}, 090501 (2006).

\bibitem{16}Generation of Dicke states using adiabatic passage. Toyoda, K. et al. \textit{Phy. Rev. A} 83, 022315 (2011).

\bibitem{17}Measurement of the coherence time of the ground-state Zeeman states in $^{40}$Ca$^+$. Haze, S., Ohno, T., Toyoda, K. \& Urabe, S. \textit{Appl. Phys. B} DOI: 10.1007/s00340-011-4603-3 (2011).
 
\end{thebibliography}
\end{document}